\documentclass[reprint,NumberedRefs]{JASA}

\usepackage{graphicx}
\usepackage{dcolumn}
\usepackage{bm}
\usepackage{amsmath,amssymb}
\usepackage{graphicx}
 \usepackage{color}
\usepackage[binary-units=true]{siunitx}

\newcommand{\ii}{\textrm{i}}
\newcommand{\ee}{\textrm{e}}

\newcommand{\dd}{\textrm{d}}

\DeclareMathOperator{\re}{Re}

\renewcommand{\vec}[1]{\bm{#1}}

\begin{document}


\title[Radiation force and torque in a cylindrical chamber]
{Acoustic radiation force and torque on  spheroidal particles in an ideal cylindrical chamber}

\author{Jos\'e P. Le\~ao-Neto}
\affiliation{Campus Arapiraca/Unidade de Ensino Penedo, Universidade Federal de Alagoas, Penedo, AL
	57200-000, Brazil}
\author{Mauricio Hoyos} 
\author{Jean-Luc Aider} 
\affiliation{%
Laboratoire de Physique et M\'ecanique des
Milieux H\'et\'erog\`enes, UMR7636 CNRS, UMPC, ESPCI, 10 rue Vauquelin, 75005 Paris, France
}%
\author{Glauber T. Silva}\email{gtomaz@fis.ufal.br}
\affiliation{%
	Physical Acoustics Group,
	Instituto de F\'isica,
	Universidade Federal de Alagoas, 
	Macei\'o, AL 57072-970, Brazil}%


\date{\today}

\begin{abstract}
We theoretically investigate how the acoustic radiation force and torque arise on a small spheroidal particle immersed in a nonviscous fluid inside an ideal cylindrical chamber. 
The ideal chamber comprises a hard top and bottom (rigid boundary condition), and a soft or hard lateral wall. 
By assuming the particle is much smaller than the acoustic wavelength, we present analytical expressions of the radiation force and torque caused by an acoustic wave of arbitrary shape.
Unlike previous results, these expressions are given relative to a fixed laboratory frame.
Our model is showcased for analyzing the behavior of an elongated metallic microspheroid (with a $10:1$ aspect ratio)
in a half-wavelength acoustofluidic chamber with a few millimeters diameter.
The results show the radiation torque aligns the microspheroid along the nodal plane, and the radiation force causes a translational motion with a speed of up to one body length per second.
At last, we discuss the implications of this study to propelled nanorods by ultrasound.
\end{abstract}


\maketitle

\section{Introduction}
Techniques for particle  manipulation  in acoustofluidic chambers 
(acoustic resonators at millimeter-scale and smaller) have been extensively used in cell separation and sorting,\cite{Ozcelik2018} microparticle patterning,\cite{Silva2019} and vesicle deformation.\cite{Mishra2014,Silva2019a}
At the core of these methods is the radiation force of acoustic waves.
This phenomenon is a stationary force caused by the linear-momentum flux change during the scattering of an incoming acoustic wave by a particle.~\cite{Torr1984,Pessoa2020}
Another related effect is the acoustic radiation torque caused by the angular-momentum flux change due to the presence of an anisotropic or absorptive particle.\cite{Hefner1999,Anhauser2012,Zhang2011a,Silva2012,Silva2014,Toftul2019}

Computing the radiation force and torque  in acoustofluidic settings is essential 
to developing applications for cell analysis and analytical chemistry.\cite{Baudoin2019b}
On that matter,
the forces and torques caused by a standing-wave field  have been investigated considering spherical particles only.\cite{Barmatz1985,Groschl1998,Goddard2005,Hagster2007,Zhuk2012,Leao-Neto2016,Lopes2016,Xu2019}
There is
an increasing interest in studying the behavior of elongated particles
 in acoustofluidic resonators such
 as fibers,\cite{Brodeur1990,Yamahira2000} microrods,\cite{Saito1998,Schwarz2015}
 nanorods,\cite{Wang2012,Ahmed2014}
 \textit{C. elegans},\cite{Ahmed2016}
 and
 \textit{E. coli}.\cite{Gutierrez-Ramos2018}
 
Geometrically speaking, an elongated particle can be modeled as a prolate spheroid with a high aspect ratio.
The analytical solution
of the radiation force and torque exerted on a prolate spheroid by a standing plane wave has been recently derived.\cite{Marston2006a,Silva2018,Fan2008,Silva2020}
In that sense, the effects of particle compressibility and density have been accounted for by using a method based on the Born approximation.\cite{Jerome2019a,Jerome2019} 
Also, the acoustic spin-torque transfer to a spheroid has also been studied.\cite{Lopes2020}  
Another resort to compute acoustic forces and torques on complex-shaped particles rely on  numerical methods.\cite{Glynne-Jones2013,Hahn2015,Wijaya2015}
It is worth mentioning that the well-known $T$-matrix approach has also been applied to compute these fields.\cite{Gong2019,Gong2019a}

In this article, we present a theoretical model to calculate the radiation force and torque on spheroidal particles in an ideal acoustic chamber
filled with a nonviscous fluid.
Our approach is based on the exact expressions of these fields
to the dipole approximation as obtained in Ref.~\onlinecite{Lima2020}.
We transform the radiation force and torque expressions to a fixed laborat
ory frame 
in which  the particle dynamics can be analyzed.
Thus, we focus our investigation on a chamber that produces a single levitation plane (half-wavelength trapping device) with radially symmetric modes.
This appears to be more suitable for studying living matter\cite{Gutierrez-Ramos2018} and developing techniques of cell culture.~\cite{Tait2019}

We apply the developed model to study artificial microswimmers (micro/nanorods) propelled by ultrasound within a cylindrical chamber. 
The synthetic microswimmers
have attracted attention due to
their potential use for drug delivery\cite{Garcia-Gradilla2014} and activation inside living cells.\cite{Wang2014}
However,
the propulsion mechanism of microswimmers propelled by ultrasound 
is still a matter of debate.
Nadal and Lauga\cite{Nadal2014} proposed an acoustic streaming model 
based on the asymmetry of a near-spherical particle that is vibrating at the wave frequency.
Collins \textit{et al.}\cite{Collis2017} included density asymmetry to this model. 
However, a recent article questioned the validity of the acoustic streaming model for a vibrating near-sphere  at low Reynolds number.\cite{Lippera2019}
In our model, we consider an artificial microswimmer as a slender microspheroid.
We   predict the microswimmer is trapped in a levitation plane, not necessarily  a nodal plane, due to the axial radiation force.
When the levitation and nodal planes coincide, the radiation torque aligns the microspheroid perpendicularly to the chamber's principal axis.
The radial radiation force causes an in-plane particle movement with a
 speed of about one body length per second (BL\,s$^{-1}$).
This suggests the radiation force minimally contributes to the observed fast speeds of microswimmers, e.g., up to $70$\,BL~s$^{-1}$.\cite{Wang2012}
 Although our model does not explain microswimmers' propulsion mechanism,
 it presents some useful insights into the dynamics of these objects in a cylindrical chamber.

\section{Physical model}

\subsection{Acoustic equations}
The  interaction between an acoustic wave and a particle takes place inside a cylindrical chamber filled with a liquid of density $\rho_0$, adiabatic speed of sound $c_0$, and compressibility $\beta_0=1/\rho_0 c_0^2$.
The chamber has radius $R$ and height $H$.
The acoustic excitation has angular frequency $\omega$, with  corresponding  wavenumber $k=\omega/c_0=2\pi/\lambda$, where  $\lambda$ is the acoustic wavelength.
We use the complex-phase representation to express the acoustic pressure and fluid velocity,  $p(\vec{r},t)=p(\vec{r})\ee^{-\ii \omega t}$ and 
$\vec{v}(\vec{r},t)=\vec{v}(\vec{r})\ee^{-\ii \omega t}$,
respectively.
Here $\ii$ is the imaginary unit, $\vec{r}$ is position vector,
and $t$ is time.

The wave dynamics in a nonviscous fluid is described by the well-known acoustic equations
\begin{subequations}
	\label{dynamic_eqs}
	\begin{align}
	&\left(\nabla^2 + k^2\right) p = 0,
	\label{helm_pressure}\\
		\label{euler}
	&\vec{v} =  \frac{\nabla p}{\ii\rho_0c_0 k}.
	\end{align}
\end{subequations}
The  term $\ee^{-\ii \omega t}$ is omitted for readability.
The acoustic equations are complemented by  boundary condition at the top, bottom, and walls of the cavity.  
\begin{figure}
    \centering
    \includegraphics[scale=.7]{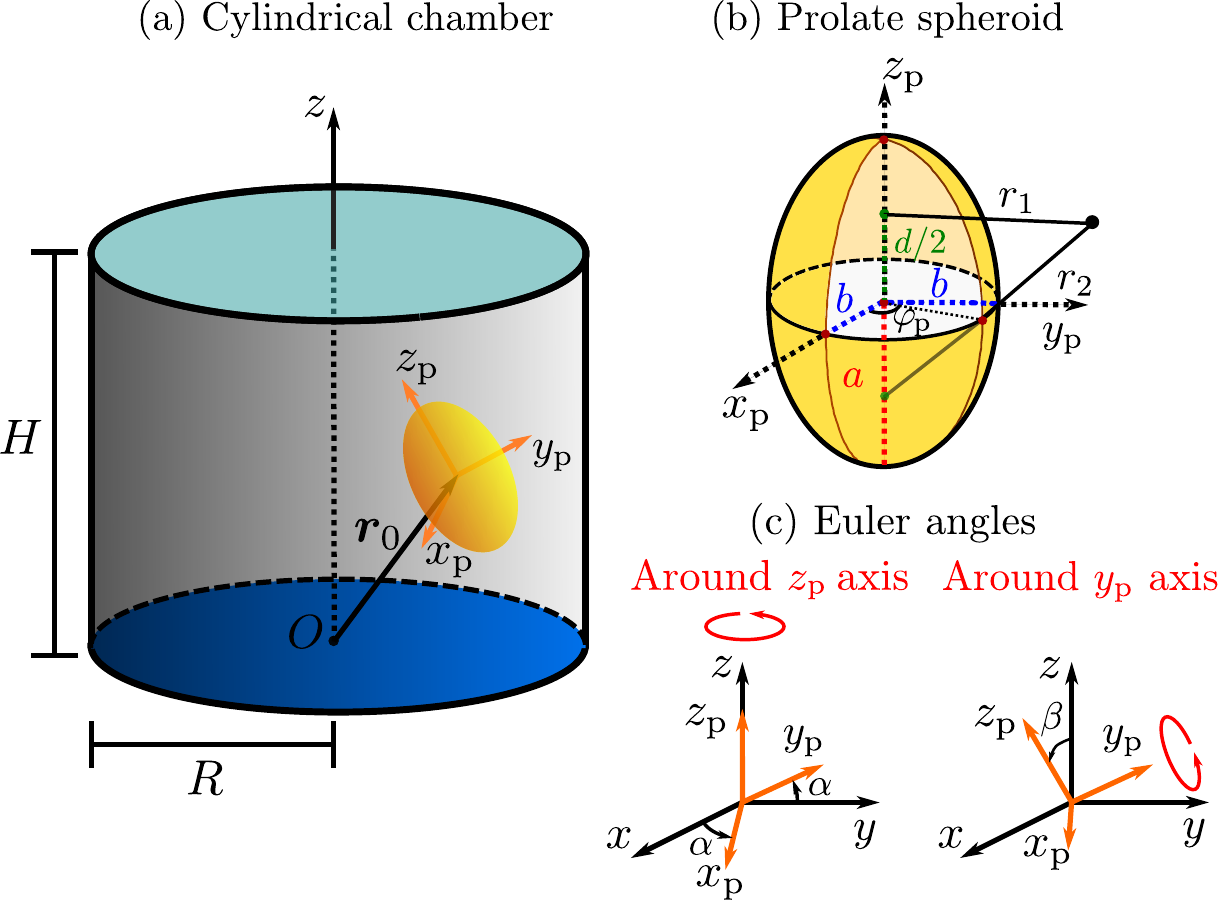}
    \caption{(a) The cylindrical acoustic chamber with a (yellow) spheroid located at $\vec{r}_0$ regarding the laboratory frame $O$ in the center of the chamber's bottom.
    (b) The prolate spheroid with major and minor semiaxis denoted by $a$ and $b$, respectively. 
    The interfocal distance is $d$.
    The quantities $r_1$ and $r_2$ are the distance from the foci to a field point.
    (c) The rotational transformations through the Euler angles $\alpha$ and $\beta$, which take the laboratory $(x,y,z)$ to  particle frame $(x_\text{p},y_\text{p},z_\text{p})$.
    \label{fig:problem}
    }
\end{figure}

\subsection{Prolate spheroidal particle}
We assume the interacting particle with the acoustic wave is a prolate spheroid, which is generated by rotating an ellipse around its major axis.
Let us define the particle frame of reference as a right-handed  system $O_\text{p}(x_\text{p},y_\text{p},z_\text{p})$ placed in
the geometric center of the spheroid.
The corresponding unit vectors of the system are 
$\vec{e}_{x_\text{p}}$, $\vec{e}_{y_\text{p}}$, and $\vec{e}_{z_\text{p}}$.
The spheroid foci are at
$(0,0,\pm d/2)$, with $r_1$ and $r_2$
being the distance from the foci to a field point--see Fig.~\ref{fig:problem}.
The prolate spheroidal coordinates $(\xi_\text{p},\eta_\text{p},\varphi_\text{p})$ are defined by
\begin{subequations}
    \begin{align}
        \xi_\text{p}& = \frac{r_1+r_2}{d}, \quad \xi_\text{p} \ge 1,\\
        \eta_\text{p} & = \frac{r_1 - r_2}{d}, \quad -1\le \eta_\text{p}\le 1,\\
        \varphi_\text{p} &= \tan^{-1}\left(\frac{y_\text{p}}{x_\text{p}}\right), \quad
        0\le \varphi_\text{p}<2\pi,
    \end{align}
\end{subequations}
with the isosurface
\begin{equation}
\label{xi0}
\xi_\text{p}=\xi_0= \frac{1}{\sqrt{1 - (\frac{b}{a})^2}}    
\end{equation}
corresponding to the particle surface.
Also, the particle major and minor axis are denoted by  $2a$ and $2b$, respectively.
While the interfocal distance and particle volume are given, respectively, by  
$d=2\sqrt{a^2-b^2}$ and $V_\text{p}=4\pi a b^2/3$.
The spheroid orientation in the particle frame coincides to the $z_\text{p}$ axis,
$\vec{d}_\text{p}= d \vec{e}_{z_\text{p}}$.

Note that
a sphere of radius $a$ is recovered by setting
$d\rightarrow 0$, $\xi_{0}\rightarrow \infty,$ and $\xi_{0} d/2 \rightarrow a$.
Whereas, a slender spheroid corresponds to the limit $\xi_0\rightarrow 1$ with a constant $d$.
In contrast, a slender spheroid results from $\xi_0\sim 1$.

\subsection{Particle versus laboratory frame of reference}

It is convenient to describe the wave-particle interaction in an inertial frame   $O(x,y,z)$ referred to as the laboratory system.
In Fig.~\ref{fig:problem}(a), we see the origin of the laboratory frame is positioned at the center of the chamber's bottom. 
And the particle position  is denoted by vector $\vec{r}_0$.
Since the spheroidal particle is invariant under rotations around its  major axis, we need only two Euler angles $(\alpha,\beta)$ to transform one frame to the other--see Fig.~\ref{fig:problem}(c).
The transformation  from the laboratory to  particle frame
is constructed as follows.
A positive rotation of an azimuthal angle $\alpha$ around the $z_\text{p}$ axis is followed by a rotation of a polar angle $\beta$ about the new $y_\text{p}$ axis.
By a positive rotation we mean a counterclockwise rotation as seen from the top of the rotation axis. 
The particle orientation in the laboratory frame is
then given by
\begin{subequations}
    \begin{align}
        \nonumber
        \vec{d} &= \mathbf{R}(\alpha,\beta)\vec{d}_\text{p}
        \\ &= d \left(\cos \alpha\sin  \beta \, \vec{e}_x +
        \sin \alpha \sin \beta \, \vec{e}_y + \cos\beta\, \vec{e}_z \right),\\
        \mathbf{R} ( \alpha, \beta)&= 
        \left(
        \begin{array}{ccc}
            \cos\alpha \cos\beta & -\sin\alpha & \cos\alpha \sin\beta\\
            \sin\alpha \cos\beta &  \cos\alpha & \sin\alpha \sin\beta\\
            -\sin\beta & 0 & \cos\beta\\
        \end{array}
        \right),
        \label{rotation}
    \end{align}
\end{subequations}
with $0\le\alpha<2\pi$ and $0\le\beta\le \pi$.
It is worth noticing the gradient operator is transformed as,
\begin{equation}
    \label{nabla}
    \nabla  = \mathbf{R}( \alpha, \beta)\nabla_\text{p}|_{\vec{r}_\text{p}=\vec{r}}, \quad
    \nabla_\text{p}  = \mathbf{R}^{-1}( \alpha, \beta)\nabla|_{\vec{r}=\vec{r}_\text{p}},
\end{equation}
where $\mathbf{R}^{-1}$ represents the transformation from the laboratory to the particle frame.

\subsection{Acoustic modes in a cylindrical cavity}
The acoustic modes allowed inside the cavity are the solutions of Eq.~\eqref{helm_pressure} 
in cylindrical coordinates $\vec{r}(\varrho,\varphi,z)$.
Accordingly, 
the pressure inside the chamber is\cite{Barmatz1985}
\begin{equation}
\label{pressure}
p(\vec{r}) = p_0 J_n(k_\varrho \varrho) \cos (n \varphi + \varphi_0) \cos k_z z,
\end{equation}  
where $p_0$ is the pressure magnitude, 
$J_n$ is the $n$th-order Bessel function, $k_\varrho$ and $k_z$ are
the radial and axial wave numbers, and $\varphi_0$ is an arbitrary constant.

The radial, angular, and axial  modes
are  determined from  boundary conditions.
We consider  hard boundaries at the bottom ($z=0$) and top ($z=H$) of the chamber.
While for the lateral wall ($\varrho=R$), a hard or soft boundary is assumed.
Accordingly, the fluid velocity and pressure satisfy
\begin{subequations}
\label{BCall}
\begin{align}
    &v_z(\varrho,\varphi,0)  =0, \quad  v_z(\varrho,\varphi,H)= 0, \\
    &v_\varrho(R,\varphi,z)  = 0 \text{~(hard)}, \quad p(R,\varphi,z) = 0, \text{~(soft)}.
\end{align}    
\end{subequations}
Since we do not have a tangential boundary condition, the phase $\varphi_0$ can be arbitrarily  set to zero.
The  conditions in \eqref{BCall} imply
\begin{subequations}
\label{mode_equations}
    \begin{align}
        &\sin \left( k_z H \right)  = 0,\\ 
        &J'_n({k_\varrho R}) = 0 \text{~(hard)},\quad
        J_n({k_\varrho R}) = 0 \text{~(soft)}.
    \end{align}    
\end{subequations}
Here the primed symbol denotes ordinary differentiation.
The solutions of these equations yield the axial and radial dispersion relations,
\begin{subequations}
     \label{ks}
    \begin{align}
        k_z&= k_l = \frac{ l \pi}{H}, \\
        \label{knm}
        k_\varrho&= k_{nm}=\frac{ j_{nm}}{R} \text{ (soft)},
~~\frac{ j_{nm}'}{R} \text{ (hard)},
    \end{align} 
\end{subequations}
with $n =0,1,2,\dots$; $l,m=1,2,3,\dots$.
The $m$th positive zero of the $n$th Bessel function and its derivative are $j_{nm}$ and $j'_{nm}$, respectively.
The total wave number  is given by
\begin{equation}
\label{wavenumber}
    k = \sqrt{k_l^2 + k_{nm}^2}.
\end{equation}
We see the angular frequency $\omega= k c_0$  is quantized. 

In what follows, we analyze radially-symmetric acoustic modes that forms a half-wavelength acoustofluidic chamber, $(nml)=(0m1)$.
Hence,  
the wave numbers turn to
 \begin{equation}
    k_1=\frac{\pi}{H}, \quad k_{0m} = \frac{j_{0,m}}{R}~\text{(soft)}, ~k_{0m} =\frac{j_{1,m}}{R}~\text{(hard)}.
 \end{equation}
 We  have used the relation between the zeros of the Bessel functions $j'_{0,m}=j_{1,m}$.
In Table~\ref{Tab:zeros}, we list the first five  zeros of the zeroth- and first-order Bessel functions for reference.
\begin{table}[]
\caption{The first five zeros of the zeroth- and first-order Bessel functions.~\cite{WeissteinWeb}
\label{Tab:zeros} 
}
\begin{tabular}{lccccc}
\hline
$m$       & 1      & 2      & 3       & 4       & 5       \\
\hline \hline
$j_{0,m}$ & 2.4048 & 5.5201 & 8.6537  & 11.7915 & 14.9309 \\
$j_{1,m}$ & 3.8317 & 7.0156 & 10.1735 & 13.3237 & 16.4706\\
\hline
\end{tabular}
\end{table}

We now express the pressure of the radially-symmetric modes,
\begin{equation}		
\label{p_cyl}
 p_{0m1} =  p_0 J_0(k_{0m}\varrho)  \cos (k_1 {z}).
\end{equation}
Substituting this equation into Eq.~\eqref{euler} yields the radial and axial components of the fluid velocity 
\begin{subequations}
\label{v_cyl}
    \begin{align}
        v_\varrho &=  \frac{ \ii v_0 k_{0m} }{k } J_1(k_{0m}{\varrho})\cos (k_1 {z}),\\
        v_z &=  \frac{\ii v_0  k_1 }{k }J_0(k_{0m}{\varrho}) \sin (k_1 {z}),
        \label{vzcyl}
    \end{align}
\end{subequations}
where $v_0= p_0/\rho_0 c_0$ is the  peak velocity.
Note we have used $J_0'(x)=-J_1(x)$.

\subsection{Scale analysis}
We assume that the particle is a subwavelength spheroid much smaller than the wavelength, which corresponds to the so-called Rayleigh scattering limit.
The particle smallness is quantified through the size factor
\begin{equation}
\label{kd}
k a = \frac{2 \pi a}{\lambda} \ll 1.
\end{equation} 
Clearly, the minor semiaxis $b$ also satisfies this condition.
We also restrict our analysis to particles much smaller to the chamber, 
$    a,b \ll H, R.$

Another effect that may appear in an acoustofluidic chamber is
the acoustic streaming, which appear near boundaries.
Acoustic streaming close to the chamber walls produces causes a drag force on the particle, 
while  near the particle surface, it
can alter the radiation force\cite{Doinikov1994,Setness2012,Baasch2019} and  produce a viscous torque.\cite{Lee1989}
As a diffusive process, streaming has a  characteristic length known as 
the viscous boundary layer,
$\delta=(2\mu_0/\rho_0 \omega)^{1/2}$,
with $\mu_0$ being the dynamic viscosity of the fluid.
To avoid streaming effects, we should consider particles much larger than this parameter, 
$
    \delta\ll a,b.
$
For example, an acoustic wave of a frequency greater than $\SI{1}{\mega\hertz}$
(a typical lower limit for  acoustofluidic devices) in water 
generates a viscous boundary layer  
 $\delta<\SI{0.84}{\micro\meter}$.

\section{Wave-particle nonlinear interaction}
\subsection{Acoustic radiation force}
The radiation force imparted on a subwavelength spheroidal particle by a stationary wave is expressed by~\cite{Lima2020}
\begin{subequations}
	\label{Fgrad}
	\begin{align}
	&\vec{F}^\text{rad}_\text{p}(\vec{0}) = - \nabla_\text{p} U_\text{p}(\vec{0}),\\
	\nonumber
	&U_\text{p}=\pi a^3\\
	&\biggl[\frac{\beta_0 f_{00}}{3} |p|^2 - \frac{\rho_0}{2} \biggl(\frac{f_{11}}{2}(| v_{x_\text{p}}|^2 + | v_{y_\text{p}}|^2) + f_{10} | v_{z_\text{p}}|^2 \biggl)\biggr],
	\label{RadPot}
	\end{align}
\end{subequations}
where $E_0=\beta_0 p_0^2/2$ is the characteristic energy density, and $\vec{v}_\text{p}= (v_{x_\text{p}}, v_{y_\text{p}},v_{z_\text{p}})$ is 
the fluid velocity in the particle frame.
Considering a rigid particle, 
the scattering amplitudes of the  monopole $f_{00}$, axial  $f_{10}$ and transverse $f_{11}$ dipole  modes are given by\cite{Lima2020}
\begin{subequations}
	\label{factors}
	\begin{align}
	f_{00} &=   1 - \xi_0^{-2},\\
	f_{10} &=\frac{2}{3\xi_0^3}\left[\frac{\xi_0}{\xi_0^2-1}-\ln\left(\frac{\xi_0+1}{\sqrt{\xi_0^2-1}}\right)\right]^{-1} ,\\
	f_{11}
	&=\frac{8}{3 \xi_0^3}\left[\frac{2-\xi_0^2}{\xi_0(\xi_0^2-1)}+\ln\left(\frac{\xi_0+1}{\sqrt{\xi_0^2-1}}\right)\right]^{-1}.
	\end{align}
\end{subequations}
These factors depend on the particle aspect ratio $a/b$ through the  parameter $\xi_0$ introduced in Eq.~\eqref{xi0}.
After inspecting \eqref{factors}, we find the following inequalities 
\begin{equation}
\label{f-inequalities}
   0<f_{00}<f_{11}<2, \quad 
    0< 2f_{10}< f_{11}<2.
\end{equation} 
As the particle geometry becomes spherical, the dipole factors turn into
$f_{11}\rightarrow 2 f_{10}.$
Whereas,  slender particles scatter much less acoustic waves,
\begin{equation}
    \label{verythin}
    f_{00},f_{10},f_{11} \rightarrow 0\text{ as } {\xi_0\rightarrow 1}. 
\end{equation}

It is more convenient to analyze the radiation force on the particle in the laboratory frame.
To this end, we have to express the acoustic fields of Eq.~\eqref{RadPot} in the laboratory frame.
By inserting the velocity components of \eqref{vpcart} into \eqref{RadPot}, we obtain  the radiation force
potential  in this frame  as
\begin{align}
    \nonumber
    U=& \pi  a^3  \biggl[
        \frac{\beta_0f_{00}}{3}  |p|^2-\frac{\rho_0}{2} \biggl(
            f_{10} | (v_x \cos \alpha + v_y \sin \alpha ) \sin \beta\\
            \nonumber
            &+ v_z \cos\beta|^2 +\frac{1}{2} 
            f_{11} \bigl[
                |v_x \sin\alpha -v_y \cos \alpha|^2 \\
                &+ |v_x \cos \alpha\,\cos \beta + v_y \sin \alpha\,\cos \beta - v_z \sin \beta|^2
            \bigr]
        \biggr)
    \biggr].
    \label{U_cart}
\end{align}
To find the potential in cylindrical coordinates, we use $v_x=v_\varrho
\cos\varphi$,  $v_y=v_\varrho \sin\varphi$.
Thus,  we have
\begin{align}
    \label{U_cyl}
    \nonumber
    U =&\pi a^3 \biggl[
        \frac{\beta_0f_{00}}{3} |p|^2 - \frac{\rho_0}{2} \biggl(
            f_{10}|v_\varrho \sin \beta \cos (\alpha-\varphi) \\
            \nonumber
            &+ v_z \cos \beta  |^2  + \frac{f_{11}}{2} \bigl[
                |v_\varrho \cos\beta\, \cos(\alpha - \varphi) \\
                &- v_z \sin \beta|^2 + |v_\varrho|^2 
                \sin^2 (\alpha -\varphi)
            \bigr]
        \biggr)
    \biggr].
\end{align}
Now, substituting the pressure and fluid velocity components given in Eqs.~\eqref{p_cyl} and \eqref{v_cyl}
into Eq.~\eqref{U_cyl}, we obtain the potential of the  radially-symmetric acoustic modes,
\begin{widetext}
    \begin{align}
        \nonumber
        U_{0m1} &= U_0 \biggl\{
            \frac{2 f_{00}}{3} \cos^2(k_1 {z}) J_0^2(k_{0m}{\varrho}) 
            - f_{10} \biggl[\frac{k_1}{k} \sin (k_1{z})   J_0(k_{0m}{\varrho})\cos \beta
            + \frac{k_{0m}}{k} \cos (k_1{z}) J_1(k_{0m}{\varrho})\,\cos (\alpha-\varphi)\, \sin\beta \biggr]^2\\
            \nonumber
            &- \frac{f_{11}}{2}\biggl[
                \biggl( 
                    \frac{k_1}{k} \sin(k_1{z}) J_0(k_{0m}{\varrho}) \sin \beta
                    - \frac{k_{0m}}{k} \cos(k_1{z}) J_1(k_{0m}{\varrho}) \cos \beta\, \cos(\alpha-\varphi) 
                \biggr)^2 
                + \left(\frac{k_{0m}}{k}\right)^2 \cos^2 (k_1 z) J_1^2(k_{0m} \varrho) \\  
                &\sin^2(\alpha-\varphi)
           \biggr] 
        \biggr\},
        \label{U0mmz}
    \end{align}
\end{widetext}
where $U_0=\pi  a^3  E_0$ is the peak potential.
For simplicity, we drop the sub-index 0 of the particle position in cylindrical coordinates,   $\vec{r}_0=(\varrho,\varphi,z)$.

By fixing the height and diameter of the chamber, the normalized potential $\tilde{U}_{0m1}= U_{0m1}/U_{0}$ depends only on the particle aspect ratio $a/b$
through the scattering factors $f_{00}$, $f_{10}$, and $f_{11}$.
The potential also depends on the
orientation  angles $\alpha$ and $\beta$, and to
 the azimuthal angle $\varphi$, albeit
the $(0m1)$ acoustic mode in Eq.~\eqref{p_cyl} has circular symmetry.
As the  particle  becomes spherical ($f_{11}\rightarrow 2f_{10}$), Eq.~\eqref{U0mmz} reduces to the radiation potential of a spherical particle as given in Ref.~\onlinecite[Eq.~1]{Barmatz1985}, with $m=0$ in the reference's notation.

Having discussed how  the  potential function is obtained, 
we  are able to derive the radiation force  in the laboratory frame. 
From Eqs.~\eqref{rotation} and \eqref{nabla}, we find this force as minus the gradient of the potential  given in Eq.~\eqref{U0mmz}, 
\begin{align}
\nonumber
    \vec{F}^\text{rad} &= \mathbf{R}^{-1}(\alpha,\beta) \vec{F}^\text{rad}_\text{p}
    = -\mathbf{R}^{-1}(\alpha,\beta)\nabla_\text{p} U_\text{p}(\vec{0})\\
    &=-\nabla U(\vec{r}_0).
    \label{RFLab}
\end{align}

Thus far, we derived the exact solution of the radiation force problem 
for the particle  placed anywhere inside the chamber.
We can distill this solution for two particular cases, namely,  along the chamber's axis of symmetry and at the nodal plane.
For the first case, the potential and radiation force are derived using Eqs.~\eqref{p_cyl} and \eqref{v_cyl} into Eq.~\eqref{U0mmz} and setting $\varrho = 0$. 
The obtained result is used in Eq.~\eqref{RFLab}.
Accordingly, we arrive at
\begin{subequations}
\begin{align}
 \nonumber
 {U}_{0m1} &=
     \frac{U_0}{6 } \biggl[
     4 f_{00}  \cos^2(k_1 z) - 3 \left(\frac{k_1}{k}\right)^2 \sin^2(k_1 z)\\ 
     &(2 f_{10} \cos^2 \beta + f_{11} \sin^2 \beta)\biggr],\\
     \label{Fz}
      F_z^\text{rad}&= F_{0,z} \Phi_\text{a}  \sin (2k_1 z), \\
    \Phi_\text{a}&= \frac{2 f_{00}}{3} +
    \left(\frac{k_1}{k}\right)^2
    \left( f_{10} \cos^2\beta + \frac{f_{11}}{2}\sin^2\beta \right),
 \end{align}
 \end{subequations}
 with $F_{0,z} = k_1 U_0$ being the axial force magnitude.
 The function $\Phi_\text{a}$ is the 
 axial acoustophoretic factor which  depends on 
 the scattering modes and orientation angle $\beta$.
Referring to the inequalities in \eqref{f-inequalities}, we conclude that
$\Phi_\text{a}>0$. 
When effects of gravity can be neglected,
the rigid spheroidal particle is  trapped in the pressure node, $z_\text{eq}=H/2$.
Note  the maximum axial force corresponds to $F_{z,\text{max}}=F_{0,z}\Phi_a$ at $z=H/4, 3H/4$.

To obtain the  radiation force  potential in the  nodal plane $z_\text{eq}=H/2$,
we see from~\eqref{v_cyl}  the pressure and the radial component of  the fluid velocity vanish, $p_{0m1}=0$
and $v_\varrho = 0$.
From Eq.~\eqref{U_cyl}, we find
\begin{subequations}
\begin{align}
\label{U_cyl2}
    U_{0m1} &=-\pi a^3 \Phi_\text{r} (\beta) \frac{\rho_0| v_z|^2}{2},\\
    \Phi_\text{r} (\beta)&= 
    f_{10} \cos^2 \beta + \frac{f_{11}}{2} \sin^2\beta.
\label{Phir}
\end{align}
\end{subequations}
The radiation force potential is  a function of the axial component of the kinetic energy density.
Besides,  the  acoustophoretic factor $\Phi_\text{r}$
does not depend on  the monopole scattering mode $f_{00}$.
This happens because the pressure vanishes at the nodal plane and so does the monopole term in Eq.~\eqref{U0mmz}.
After substituting Eq.~\eqref{vzcyl} into Eq.~\eqref{U_cyl2} and replacing the result into 
Eq.~\eqref{RFLab}, we obtain
the potential and radial radiation force as
\begin{subequations}
\begin{align}
\label{Uradial0}
    {U}_{0m1} &= 
        -\left(\frac{k_1}{k}\right)^2 U_0\Phi_\text{r}(\beta)J^2_0(k_{0,m} \varrho),\\
    \label{Fradial}
    F^\text{rad}_\varrho&= 
        -F_{0,\varrho}    \Phi_\text{r} (\beta) J_0(k_{0m} \varrho) J_1(k_{0m} \varrho),\\
    F_{0,\varrho} &= 
        2 \left(\frac{k_1}{k}\right)^2 k_{0m}  U_0,
\end{align}
\end{subequations}
with $F_{0,\varrho}$
being the force magnitude.
The radial acoustic traps correspond to the the minima of the potential function, while
the largest force occurs at $k_{0m} \varrho = 1.081$,
with corresponding magnitude of
$    F_{\varrho,\text{max}}^\text{rad}= 0.338 F_{0,\varrho} \Phi_\text{r}.$
For a rigid particle, the radial acoustophoretic factor  is  positive and the potential minima are obtained by solving the equation  ${J_0^2}'(k_{0m}\varrho)=0$.
This corresponds to find the zeros of the first-order Bessel function. 
Hence, the position of the $i$th radial trapping point is at 
\begin{equation}
    \varrho_{i,m} = \frac{j_{1,i-1}}{j_{0,m}} R~\text{(soft)},
    ~\frac{j_{1,i-1}}{j_{1,m}} R~\text{(hard)}, \quad m=1,2,\dots
\end{equation}
Here we consider $j_{1,0}=0$.
The primary trap corresponds to $\varrho_{1,m} = 0$ regardless the lateral boundary condition, e.g., soft or hard wall.
To determine  the  second trap position, we refer  to  Table~\ref{Tab:zeros},  
$\varrho_{2,1} = 0.63 R$ (soft wall)
and $\varrho_{2,1} = R$ (hard  wall).
We see  soft walled chambers are able to produce only a middle  trap.
Whereas, the second trap of a hard walled chamber is located at 
the lateral wall.

\subsection{Acoustic radiation torque}
\label{Sec:torque}
The acoustic radiation torque exerted on the spheroidal particle by the  acoustic mode described in Eq.~\eqref{p_cyl}, is given in the particle frame by\cite{Lima2020}
\begin{subequations}
\begin{align}
\vec{\tau}^\text{rad}_\text{p}
&= -\pi a^3    \chi 
\left(\vec{e}_{z_\text{p}}\times {\bf P}_\text{p}\cdot  \vec{e}_{z_\text{p}}\right)_{\vec{r}_\text{p}=\vec{0}},
\label{torque_rayleigh4}\\
{\bf P}_\text{p}&=\frac{\rho_0}{2} \re[\vec{v}_\text{p}\vec{v}_\text{p}^*]=
\frac{\rho_0}{2} \re \bigl[
    v_i v_j^* \vec{e}_i\vec{e}_j
    \bigr],  ~~i,j= x_\text{p}, y_\text{p}, z_\text{p},
\label{LMFlux}
\end{align}
\end{subequations}
where $\chi=f_{11}- 2 f_{10}>0 $ is the  gyroacoustic factor and
${\bf P}_\text{p}$ is the time-average of 
the linear momentum flux (a second-rank tensor) relative to the particle frame.
We express the projection of the linear momentum flux onto the axial direction as
$\mathbf{P}_\text{p}\cdot \vec{e}_{z_\text{p}} = (\rho_0/2) \re[v_{z_\text{p}}^* \vec{v}_{\text{p}}]$.
Carrying on  the calculations, we arrive at
\begin{equation}
\vec{\tau}^\text{rad}_\text{p}
= \frac{\pi a^3}{2} \chi\, \rho_0 \re\left[
v_{y_\text{p}}v_{z_\text{p}}^*\vec{e}_{x_\text{p}} - v_{x_\text{p}}v_{z_\text{p}}^*\vec{e}_{y_\text{p}} \right].
\label{torque_rayleigh2}	
\end{equation}
To find the radiation torque in the laboratory frame, we apply the  rotation matrix $\mathbf{R}$ into  Eq.~\eqref{torque_rayleigh2},
\begin{align}
	\label{torque_lab}
    \nonumber
    \vec{\tau}^\text{rad}&=\mathbf{R}(\alpha,\beta) \vec{\tau}^\text{rad}_\text{p}\\
    \nonumber
    & = \frac{\pi a^3}{2}\chi\rho_0 \re
    \bigl[
 (v_{y_\text{p}}v_{z_\text{p}}^*\cos \alpha \cos \beta + v_{x_\text{p}}v_{z_\text{p}}^* \sin \alpha )\vec{e}_x\\
 \nonumber
 &+ (v_{y_\text{p}}v_{z_\text{p}}^* \sin \alpha \cos \beta - v_{x_\text{p}}v_{z_\text{p}}^* \cos \alpha) \vec{e}_y \\
 &- v_{y_\text{p}}v_{z_\text{p}}^*  \sin\beta\,
 \vec{e}_z
 \bigr].
\end{align}
Substituting the fluid velocity components given in Eq.~\eqref{v_cyl} into Eq.~\eqref{vpcyl} and
replacing the result into Eq.~\eqref{torque_lab}, we obtain
\begin{widetext}
\begin{subequations}
\label{torque_all}
	\begin{align}
	\nonumber
	\tau_x &=
	-\frac{\pi a^3  \chi E_0}{2} \biggl[\left(\frac{k_1}{k}\right)^2 \sin 2\beta \,\sin \alpha\,\sin^2(k_1 {z})\, J_0^2(k_{0m}{\varrho})
	+ \frac{k_1 k_{0m}}{k^2} \sin (2k_1{z})\, J_0(k_{0m}{\varrho})J_1(k_{0m}{\varrho})\\
	&[\sin^2\beta\,\sin\alpha
	 \cos(\alpha-\varphi) +\cos^2 \beta\,\sin\varphi] - \left(\frac{k_{0m}}{k}\right)^2\sin 2\beta \,
	\cos^2 (k_1 z)\,\sin\varphi\ J_1^2(k_{0m}{\varrho})\cos(\alpha-\varphi)\,\biggr],\\ 
		\nonumber
		{\tau}_{y}& =  
		\frac{\pi a^3  \chi E_0}{2} \biggl[\left(\frac{k_1}{k}\right)^2 \sin 2\beta\,\cos \alpha\,\sin^2(k_1 {z})\, J_0^2(k_{0m}{\varrho})
	+ \frac{k_1 k_{0m}}{k^2} \sin (2k_1{z})\, J_0(k_{0m}{\varrho})J_1(k_{0m}{\varrho})\\
	&[\sin^2\beta\,\cos\alpha  \cos(\alpha-\varphi)-\cos^2\beta\,\cos\varphi]
	- \left(\frac{k_{0m}}{k}\right)^2\sin (2\beta)\,
	\cos^2 (k_1{z})\, J_1^2(k_{0m}{\varrho})\cos(\alpha-\varphi)\, \cos \varphi\biggr],\\ 
		\nonumber
		{\tau}_{z}& =  
		\frac{\pi a^3  \chi E_0}{2} \biggl[\frac{k_1 k_{0m}}{2k^2}  
		\sin 2\beta\,\sin(\alpha-\varphi)\,\sin(2 k_1{z})J_0(k_{0m}{\varrho})J_1(k_{0m}{\varrho})+\left(\frac{k_{0m}}{k}\right)^2\sin^2\beta\,
	\cos^2 (k_1{z})\sin[2(\alpha-\varphi)]\\
	&  J_1^2(k_{0m}{\varrho})\biggr].
		\end{align}
\end{subequations}
\end{widetext}

When the particle is trapped at $z_\text{eq}=H/2$,
 we see from~\eqref{v_cyl} that the radial component of  the fluid velocity vanishes $v_\varrho = 0$.
Hence, referring to Eqs.~\eqref{vpcyl} and \eqref{torque_lab}, the radiation torque reduces to
\begin{equation}
\label{torque_zH}
    \vec{\tau}^\text{rad}
    = \pi a^3 \chi \sin 2\beta \,\frac{\rho_0 |v_z|^2}{4}\,\vec{e}_\alpha.
\end{equation}
The unit vector $\vec{e}_\alpha= \cos \alpha\, \vec{e}_y- \sin\alpha \,\vec{e}_x$
lies along the minor semiaxis pointing to the counterclockwise direction in the $xy$ plane.
The radiation torque  is proportional to the axial component of the kinetic energy density averaged in time $\rho_0 |v_z|^2/4$.
It also depends on the  orientation factor $\sin 2\beta$.
The particle is set to  rotate around the minor axis, since $\vec{e}_\alpha \cdot \vec{e}_z = 0$.
Now we replace $v_z$ in Eq.~\eqref{torque_zH} by
 Eq.~\eqref{vzcyl}  to encounter 
\begin{equation}
    \label{torque_analytic}
	\vec{\tau}^\text{rad}(\beta)
		 = \tau_0 \chi  J_0^2(k_{0m}{\varrho}) \sin2\beta\,\vec{e}_\alpha,
\end{equation}
where
$\tau_0 = \pi a^3 E_0 k_1^2/2k^2$ is the characteristic torque.
The maximum torque $\tau_\text{max}^\text{rad}=\tau_0 \chi$, which occurs at $\beta=\pi/4$ and $\varrho=0$.
The equilibrium angular position corresponds to $\beta=\pi/2$.

\begin{table}
	\caption{\label{tab:parameters}
		The physical and geometric parameters of the microswimmer in a submillimeter cylindrical chamber at room temperature and pressure.
	}
	{
		\begin{tabular}{lr}
			\hline
			\textbf{Parameter} & \textbf{Value}\\
			\hline
			\hline
			\textbf{Microspheroid (Au)}& \\
			Major semiaxis ($a$)   &  
			$\SI{10}{\micro\meter}$\\
			Minor semiaxis ($b$) & $\SI{1}{\micro\meter}$\\
			Aspect ratio ($a/b$) & 10:1\\
			Radial parameter ($\xi_0$)  &  $1.0050$\\
			Volume ($V_\text{p}$) & $\SI{41.9}{\micro\meter^3}$\\
			Density ($\rho_\text{p}$) & $\SI{19300}{\kilogram\per\meter \cubed}$\\
			Moment of inertia ($I$) & $\SI{16.3}{\nano\gram \micro\meter\squared}$\\
			Monopole mode ($f_{00}$) & $0.01$\\
			Axial dipole mode ($f_{10}$)
			& $0.0068$\\
			Transverse dipole mode ($f_{11}$)
			& $0.0261$\\
			\hline
			\textbf{Water} &\\
			Density ($\rho_0$) &  $ \SI{1000}{\kilogram \per \meter\cubed}$\\
			Speed of sound ($c_0$) & $ \SI{1492}{\meter \per \second}$\\
			\hline
			\textbf{Cylindrical chamber}\cite{Wang2012} \\
			Height ($H$) & $\SI{180}{\micro\meter}$\\
	        Radius ($R$) & 
	        $\SI{2.5}{\milli\meter}$\\
			Levitation plane $(z_\text{eq})$ & $\SI{76.5}{\micro\meter}$\\
			Energy density ($E_0$) & $\SI{15.3}{\joule\per\meter\cubed}$\\
			\hline
		\end{tabular}
	}
\end{table}


\subsection{Effects of gravity}
An actual particle of density $\rho_\text{p}$ is subjected to effects of gravity, which changes  its axial equilibrium position.
The new position  can be determined from the force
equilibrium equation
$F^\text{rad}(0,z_\text{eq}) - (\rho_\text{p} -\rho_0) V_\text{p} g = 0$,
with $g$ being the gravity acceleration.
Thus from Eq.~\eqref{Fz},
the axial equilibrium position is 
\begin{equation}
    \label{zeq}
    z_\text{eq}= \frac{H}{2} -  \frac{H}{2\pi}
   \arcsin \left[\frac{4(\rho_\text{p} - \rho_0) g H}{3\pi  \Phi_\text{a} E_0 }\left(\frac{b}{a}\right)^{2}\right].
\end{equation}
To bring the particle close to the nodal plane, we need to increase the acoustic energy density.
From Eq.~\eqref{zeq}, we see  the energy density needed to keep the particle in equilibrium is
\begin{equation}
    \label{energy_density}
    E_0 =
    \frac{4(\rho_\text{p} - \rho_0)  g H}
    { 3\pi \Phi_\text{a} \sin(2\pi z_\text{eq}/H)}
    \left(\frac{b}{a}\right)^2.
\end{equation}
We see that slender particles with $a\gg b$ require less energy to be  axially trapped.

\begin{table}[]
\caption{The theoretical predictions of the microspheroid at the nodal plane  considering the parameters of Table~\ref{tab:parameters}.
\label{tab:freq_stiffness}
}
\begin{tabular}{lrrrr}
\hline
 &\multicolumn{4}{c}{\textbf{Acoustic modes}} \\ 
 \cline{2-5}
\multicolumn{1}{c}{\textbf{Feature}}   & \multicolumn{2}{c}{\textbf{Soft}} & \multicolumn{2}{c}{\textbf{Hard}} \\ 
 \cline{2-5}
& \multicolumn{1}{c}{(011)} & \multicolumn{1}{c}{(021)} & \multicolumn{1}{c}{(011)} & \multicolumn{1}{c}{(021)} \\ 
\hline
\hline
Frequency {[}MHz{]}                           & $4.150$                  & $4.177$                  & $4.160$                  & $4.197$ \\
Radial  force, $F_{\varrho,\text{max}}^\text{rad}\,[\si{\pico\newton}]$ & $0.407$ & $0.921$ & $0.645$ & $1.160$\\ 
Translational velocity, $\dot{\varrho}\,[\si{\micro\meter\per\second}]$         & $8.185$                   & $18.55$                   & $12.98$                   & $23.35$                   \\
Trap time, $t_\varrho\,[\si{\second}]$        & $43.05$                   & $8.277$                   & $17.04$                   & $5.174$  \\
Radiation torque, $\tau_\text{max}^\text{rad}\,[\si{\nano\newton\micro\meter}]$ & $0.299$                   & $0.296$                   & $0.298$                   & $0.293$\\
Angular velocity, $\dot{\beta}\,[\si{\radian\per\second}]$              & $23.62$                    & $23.02$                   & $23.40$                   & $22.58$\\
Reorientation time, $t_\beta\,[\si{\milli\second}]$                           & $31.44$                   & $32.26$                   & $31.73$                   & $32.89$\\
 \hline
\end{tabular}
\end{table}

\subsection{Translational and angular  velocity of the particle}
Here we obtain the stationary translational and angular velocity achieved
by the particle at
the nodal plane  $z_\text{eq}= H/2$.
This analysis is restricted to particles at microscale in an aqueous solution.

To determine the translational  velocity, we assume the particle is
at $(\varrho,\varphi,H/2)$ and aligned  to the radial direction, $\beta=\pi/2$ and $\alpha=\varphi$. 
Hence the velocity is denoted by $\dot{\varrho}$, with  dot notation meaning time derivative.
As the particle moves, a drag force counteracts the radiation force,\cite{Deo2003}
\begin{subequations}
    \begin{align}
        \vec{F}^\text{drag} &= - 8 \pi a \mu_0  g_\text{f} \dot{\varrho}\, \vec{e}_\varrho,\\
        g_\text{f} &=
        \frac{1}{\xi_0[(\xi_0^2+1)\,\text{arccoth}\, \xi_0 -\xi_0]}.
    \end{align}
\end{subequations}
The geometric factor $g_\text{f}$ becomes $3/4$ for a spherical particle $(\xi_0\rightarrow \infty)$, which leads to the well-known Stoke's law, $F^\text{drag}_\text{sphere}=-6\pi \mu_0 a \dot{\varrho}$.

Using Eq.~\eqref{Fradial}, we find 
the  equation of motion of a particle moving along its major axis
as
\begin{equation}
\label{eqmotion}
 \ddot{\varrho}+ \frac{8\pi a \mu_0 g_\text{f}}{M }\dot{\varrho}= -
    \frac{F_{0,\varrho}\Phi_\text{r}}{M} J_0(k_{0m} \varrho) J_1(k_{0m} \varrho).
\end{equation}
where $M$ is  the particle's mass.
Considering a micrometer-sized particle in water, we see 
the viscous contribution overcomes inertia by far.
So the inertial term in Eq.~\eqref{eqmotion} can be neglected.
The equation of motion then becomes
\begin{equation}
\label{trans_vel}
    \dot{\varrho} = -\left(\frac{k_1}{k} \right)^2\frac{ k_{0m} a^2   \Phi_\text{r}}{4   g_\text{f}}\frac{E_0}{\mu_0}
    J_0(k_{0m} \varrho) J_1(k_{0m} \varrho).
\end{equation}
We conclude the translational speed increases with  the particle length squared.
We find the solution of Eq.~\eqref{trans_vel} for a particle in the vicinity of $\varrho=0$
with the initial position at
$\varrho(0) = \varrho_0$,
\begin{subequations}
    \begin{align}
        \varrho(t)&=\varrho_0\, \ee^{-t/t_\varrho},\\
        t_\varrho &= \left(\frac{2k}{k_1 k_{0m} a} \right)^2\frac{   g_\text{f} \mu_0}{  \Phi_\text{r} E_0}.
    \end{align}
\end{subequations}
Importantly, the characteristic trapping time $t_\varrho$ is of the order of seconds.

Turning now to the  angular velocity  induced by the radiation torque
of Eq.~\eqref{torque_analytic} on a particle  
at $(\varrho,\varphi, H/2)$.
As the radiation torque  depends only on the orientation angle $\beta$,  the angular velocity corresponds to the rate change of the orientation, $\dot{\beta}$.
Moreover, a drag torque arises on the particle,\cite{Silva2020}
\begin{subequations}
    \begin{align}
        \label{taudrag}
        \vec{\tau}^\text{drag} &= -
            8 \pi a^3 \mu_0  g_\text{t} \dot{\beta}\, \vec{e}_\alpha,\\
        g_\text{t} &= 
            \frac{4}{3\xi_0^3}\frac{1- 2\xi_0^2}{2\xi_0-         (1+\xi_0^2)\ln\left(\frac{\xi_0+1}{\xi_0-1}\right)}.
    \end{align}
\end{subequations}
The well-known result of the drag torque for a sphere, $\tau^\text{drag} = -8 \pi a^3 \mu_0 \dot{\beta}$,  is obtained by setting $\xi_0\rightarrow \infty$.

The rotational particle dynamics is described by the differential equation 
\begin{equation}
\label{rot_dyn}
    \ddot{\beta} + \frac{8 \pi a^3 \mu_0  g_\text{t} }{I}\dot{\beta} = \frac{\tau_0 \chi J_0^2(k_{0m}{\varrho})}{I}   \sin 2\beta,
\end{equation}
with $I=M (a^2+b^2)/5$ being the particle moment of inertia relative to  the minor axis.
Again the viscous effects overcome  inertia. 
So
the rotational equation of motion becomes
\begin{equation}
	      \dot{\beta}= \frac{\tau_0 \chi J_0^2(k_{0m}{\varrho}) \sin 2\beta}{8 \pi a^3 \mu_0  g_\text{t} },
\end{equation}
which can be solved by the method of separation of variables.
Let $\beta_0$ be the initial particle orientation.
Using the expression $\int \sin^{-1} 2\beta\, \dd \beta =\ln(\tan \beta)/2$, we find
\begin{subequations}
    \begin{align}
        \beta(t) &= \text{arccot}\left[\exp\left(-\frac{t}{t_\beta J_0^2(k_{0m}{\varrho})}\right)\cot \beta_0
        \right],\\
        t_\beta &=  \left(\frac{4 k}{k_1}\right)^2 \frac{ g_\text{t}  \mu_0  }{ \chi E_0}.
    \end{align}
\end{subequations}
The orientation angle asymptotically approaches $\beta=\pi/2$ (aligned with the nodal plane) as $t\rightarrow \infty$.
Slender particles $\chi\rightarrow 0$ need more time to reach equilibrium, as well as
particles far from the center.
The characteristic reorientation time $t_\beta$ is of the order of milliseconds.

The rotational-to-translational characteristic time ratio is about
\begin{equation}
    \frac{t_\beta}{t_\varrho} \sim (k_{0m} a)^2.
\end{equation}
This ratio is about $10^{-3}$
for typical acoustofluidic settings. 
\begin{figure}
\centering
\includegraphics[scale=.28]{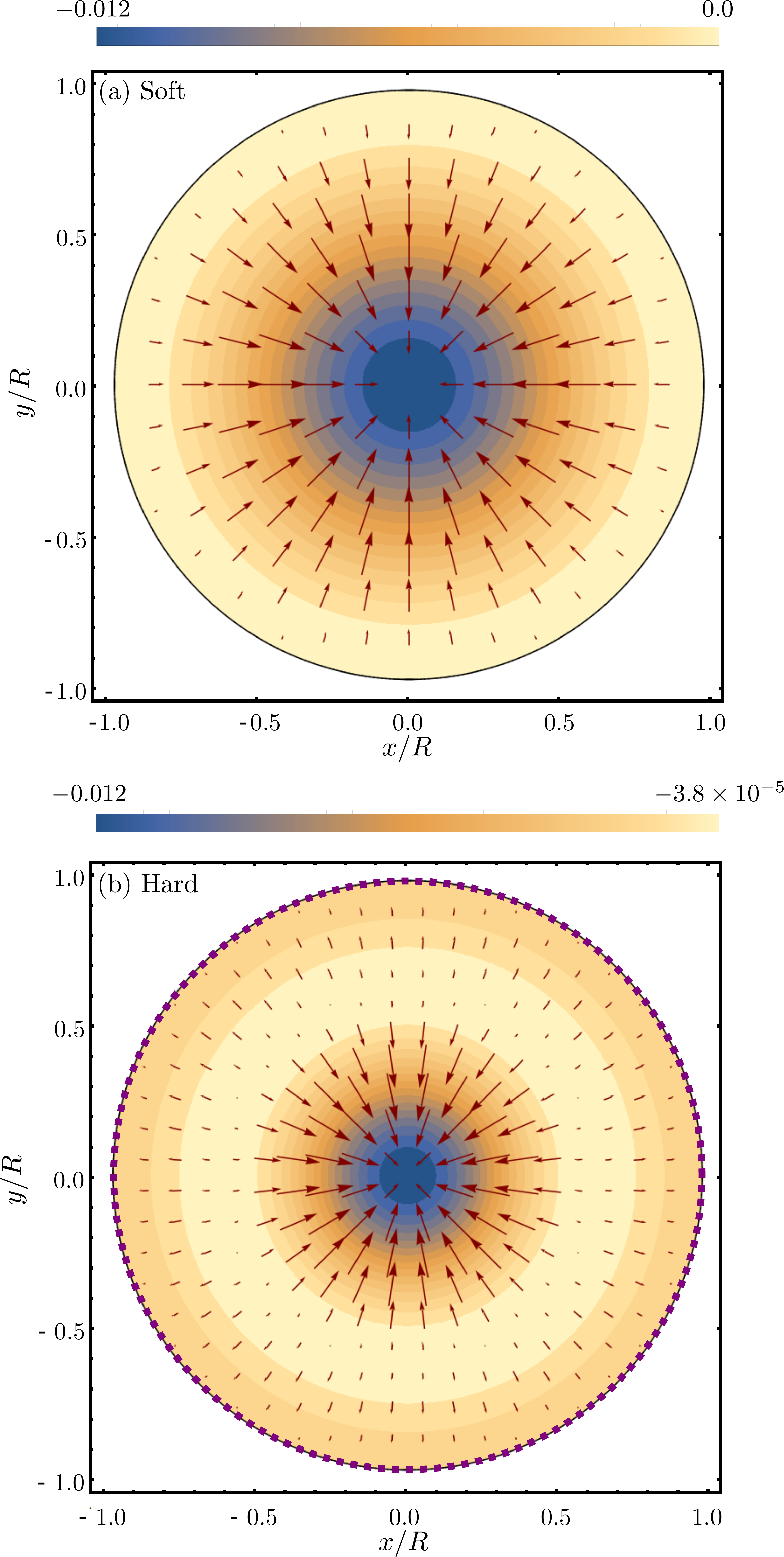} 
\caption{\label{fig:force_soft} 
The radiation force  fields (red arrows) of the microspheroid aligned to the $x$ axis. The force is generated by
the (011) acoustic mode with (a) soft  and (b) hard lateral walls.
The background contours illustrate the  potential function $U_{011}$, given by Eq.~\eqref{U0mmz},  normalized to $U_0=\SI{48.07}{\femto\joule}$.
The force fields are evaluated in the laboratory frame at the axial position $z_\text{eq} = 0.85 H/2$.
The physical parameters used here are listed in Table~\ref{tab:parameters}. 
The bluish regions correspond to the middle  trap, while the dotted-purple circle in panel (b) is the  annular trap.
}
\end{figure}

\section{Case study: {Au} microrods}
Now, we use the theory to analyze the radiation force and torque fields in a acoustofluidic chamber wherein the particles are trapped as described in Ref.~\onlinecite{Wang2012}.
In this reference,
the chamber operates at nearly $\SI{4}{\mega\hertz}$, and
the particles are metallic (Au) nanorods  with length of  few micrometers and  hundreds of nanometers wide. 
These objects can be geometrically modeled as  microspheroids with a slender shape.
As the particle width is of the order of the viscous boundary layer, we cannot applied  our method directly to these nanorods.
Nevertheless, the theory can be used to explain the behavior of wider particles with the same aspect ratio ($10:1$) of the nanorods.
In doing so, the physical parameters of our analysis are summarized in Table~\ref{tab:parameters}.
Finally,
our choice of the levitation plane position at
$z_\text{eq}=0.85 H/2$ is arbitrary, but   compatible with the previous reported levitation height  of   Au particles\cite{Dumy2019}, $0.41 H/2< z_\text{eq}<H/2$.
Hence, according to Eq.~\eqref{energy_density}, the corresponding energy density for the chosen height of the  levitation plane is $E_0=\SI{15.3}{\joule\per\meter\cubed}$.
\begin{figure}
\centering
\includegraphics[scale=.28]{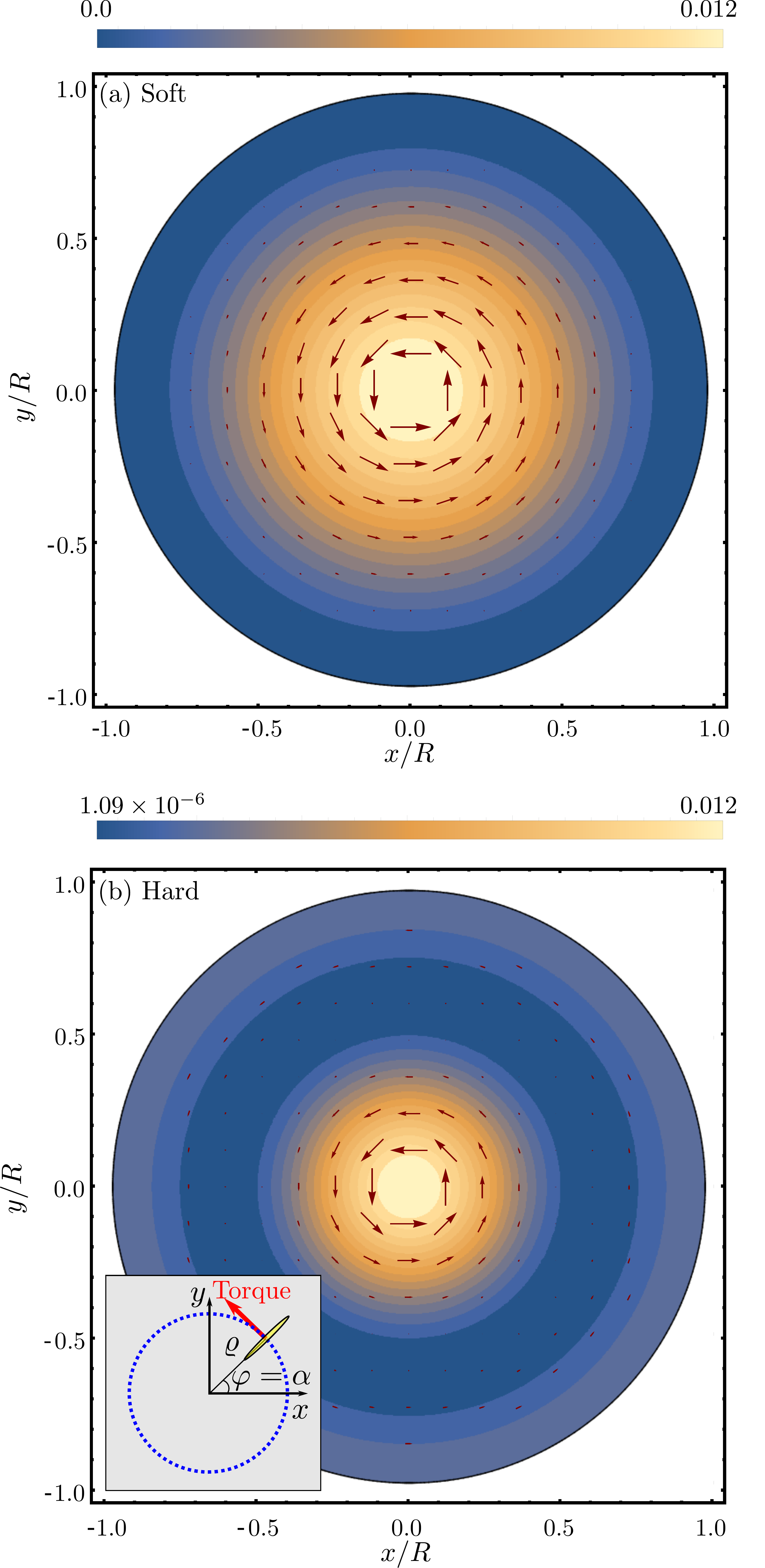} 
\caption{\label{fig:torque} 
The radiation torque  fields (red arrows) in the levitation plane at $z_\text{eq} = 0.85 H/2$ produced by
the (011) acoustic mode for (a) a soft  and (b) hard lateral wall.
The background contours illustrate the radiation torque amplitude from  Eq.~\eqref{torque_analytic} and  normalized to  $\tau_0=\SI{23.9}{\nano\newton \micro\meter}$.
The inset of panel (b) shows the microspheroid (in yellow)  at the position
$(\varrho,\varphi)$, aligned with the radial direction ($\alpha=\varphi$), and with $\beta=\pi/4$.
The radiation torque is always perpendicular to the particle orientation.
The physical parameters used here are listed in Table~\ref{tab:parameters}. 
}
\end{figure}

With all model parameters in place, we can compute some features of  the microspheroid behavior at the nodal plane for
the (011) and (021) acoustic modes. 
The results are summarized in Table~\ref{tab:freq_stiffness}.
The characteristic trap time is  of the order of seconds, and
 the reorientation time is about $\SI{31}{\milli\second}$.
Besides, the microspheroid can be as fast as one body length per second.
Note also the rigid walled chamber yields the largest radiation forces.
In contrast, the radiation torque does not change with the chamber boundary conditions at all.

In Fig.~\ref{fig:force_soft}, we show the
radiation force field (red arrows) acting on the microspheroid 
aligned with the $x$ axis
as a function
of the scaled coordinates $x/R$ and $y/R$.
The background contour plots corresponds to the force potential  $U_{011}$, which appears radially symmetric at 
 $z_\text{eq}=0.85 H/2$.
Panels (a) and (b) display the results for soft and hard lateral boundary conditions, respectively.
The bluish region corresponds to the middle trap, while the dotted-purple circle at $\varrho/R=1$ in panel (b) illustrates the annular trap. 

In Fig.~\ref{fig:torque}, we show
the radiation torque field (red arrows)  on the microspheroid 
as a function of the scaled Cartesian coordinates.
Both soft and hard wall chambers are considered with the (011) acoustic mode.
The background contour plot is the radiation torque amplitude normalized to the characteristic torque $\tau_0=\SI{48.07}{\pico\newton \micro\meter}$.
The microspheroid  position is at $(\varrho,\varphi,0.85H/2)$, with orientation along the radial direction $\alpha=\varphi$ and $\beta=\pi/4$.
We note the radiation torque has radial symmetry and points to the tangential direction $\vec{e}_\varphi$--see the inset in panel (b).
Also, a larger radiation torque is achieved in the middle area with nearly the same amplitude in both chambers.
Though the soft chamber develops a more homogeneous torque around the central area of the levitation plane.
The principal effect of the radiation torque is to reorient the particle to the angular position $\beta=\pi/2$.
In Table~\ref{tab:freq_stiffness}, we see the reorientation characteristic time is about $t_\beta = \SI{30}{\milli\second}$.
Moreover, it is independent of the lateral boundary conditions.

The particle reorientation effect was observed in millimeter-sized paper fibers caused by a standing plane wave at $\SI{72}{\kilo\hertz}$ in water.\cite{Brodeur1990}
A similar conclusion was achieved for polystyrene fibers with one-fourth of the wavelength in an acoustic resonator filled with water.\cite{Yamahira2000}
Nonetheless, an intriguing experimental observation in microgravity shows that a cluster of trapped $\SI{3}{\micro\meter}$-long nanorods in water are aligned perpendicularly to the nodal plane inside a cylindrical chamber.\cite{Dumy2020a}
On this matter, we offer the following explanation for this effect.
Firstly,  the fluid viscosity may play a significant role in the radiation torque changing the orientation equilibrium position. Secondly, with the inter-particle distances being about the  particle dimensions, the secondary radiation force becomes dominant.\cite{Silva2014b,Sepehrirahnama2015}
So one may expect the rise of secondary radiation torques. In turn, the secondary interaction torques are likely to change the particle orientation equilibrium.
Thirdly, both density and geometric asymmetries seem to have a markedly influence on the nanorods behavior.\cite{Ahmed2016}
None of these features are taken into account by our approach.


\section{Concluding remarks}
In this study, we present
analytical results of the acoustic radiation force and torque developed on a rigid (prolate) spheroidal particle inside an ideal cylindrical chamber. 
The particle is considered far smaller than the acoustic wavelength and much larger than viscous boundary layers.
The ideal chamber comprises a rigid bottom and top, with hard or soft lateral walls.
The radiation force and torque expressions are given in the laboratory frame,
paving the way to investigating the particle behavior through equations of motion.
This approach can also be used for an incident wave of arbitrary shape, as long as the beam is expressed (analytically or numerically) in Cartesian or cylindrical coordinates.

The theory is applied to calculate the radiation forces and torques acting on a microspheroid.
The model parameters are chosen to mimic the experimental setup of nanorods propelled by ultrasound.\cite{Wang2012} 
As the nonviscous approximation is assumed, 
we could not apply theory directly to the nanorods.
Notwithstanding, we keep the same aspect ratio of the nanorods ($10:1$)  but consider a 
microspheroid with a diameter of $\SI{2}{\micro\meter}$ which is larger than the  boundary layer depth.
We obtain 
the characteristic radiation force and torque, and
the particle translational and angular velocities
of the first acoustic modes of the chamber.
Furthermore,  the particles in the nodal plane are reoriented to the same direction of this plane by means of the radiation torque.
The reorientation time is of the order of milliseconds.
Whereas, the radial trap occurs after  several seconds passed.

Our model also predicts translational speeds of up to one body lengths per second (BL\,s$^{-1}$).
The speed increases with the particle length squared.
Should we applied the theory to the nanorods of Ref.~\onlinecite{Wang2012}, the speed would be at least ten times smaller.
This hints that the radial radiation force does not significantly impact the nanorods' propulsion mechanism. 

The present analysis is a solid step toward understanding the  physics behind trapping elongated particles in acoustofluidic settings.
It offers results that can be verified experimentally for systems
whose boundary conditions can be approximated to ideal conditions (hard or soft walls).
Adding thermoviscous properties of the surrounding fluid to the model is the next level
to be attained in future publications.

\begin{acknowledgments}
G. T. Silva thanks 
the Brazilian National Council for Scientific and Technological Development--CNPq (Grant number 
308357/2019-1), and Chaire Total ESPCI-Paris (2019). 
\end{acknowledgments}

\appendix
\section{Transformations of the fluid velocity field}
Going back to the transformation in Eq.~\eqref{rotation}, we see the relation between the fluid velocity in the particle and laboratory frame is expressed as
\begin{equation}
\label{vel}
  \vec{v}_\text{p}(\vec{0})= \mathbf{R}^{-1}(\alpha,\beta)\, \vec{v}(\vec{r}_0).
\end{equation}
Thus, we have in  Cartesian coordinates,
\begin{subequations}
\label{vpcart}
\begin{align}
    v_{x_\text{p}}&= 
    v_x \cos \alpha \, \cos \beta  + v_y \sin \alpha\, \cos \beta  - v_z \sin \beta,\\
    v_{y_\text{p}}&= v_y \cos \alpha - v_x \sin\alpha,\\
    v_{z_\text{p}}&= v_x \cos \alpha\, \sin \beta  + v_y \sin \alpha\, \sin \beta + v_z \cos \beta.
\end{align}
\end{subequations}
The corresponding components in
to cylindrical coordinates are obtained using $v_x=v_\varrho
\cos\varphi$,  $v_y=v_\varrho \sin\varphi$, 
\begin{subequations}
\label{vpcyl}
\begin{align}
    v_{x_\text{p}}&= 
     v_\varrho \cos \beta \cos (\alpha - \varphi) -  v_z \sin \beta,\\
    v_{y_\text{p}}&= -v_\varrho \sin(\alpha-\varphi),\\
    v_{z_\text{p}}&= v_\varrho \cos(\alpha-\varphi) \sin \beta  + v_z  \cos \beta.
\end{align}
\end{subequations}


\end{document}